\documentclass[fleqn,usenatbib]{mnras}

\usepackage{newtxtext,newtxmath}

\usepackage[T1]{fontenc}
\usepackage{ae,aecompl}

\usepackage{graphicx}	
\usepackage{amsmath}	

\usepackage{comment}
\usepackage{color}

\title [Full compressible 3D magnetohydrodynamic simulation of solar wind]{Full compressible 3D MHD simulation of solar wind }

\author[T. Matsumoto]{
Takuma MATSUMOTO \thanks{E-mail: takuma.matsumoto@nao.ac.jp}\\
National Astronomical Observatory of Japan, 2-21-1 Osawa, Mitaka, Tokyo 181-8588, Japan 
}

\date{Accepted XXX. Received YYY; in original form ZZZ}

\pubyear{2015}

\begin{document}
\label{firstpage}
\pagerange{\pageref{firstpage}--\pageref{lastpage}}
\maketitle

\begin{abstract}
Identifying the heating mechanisms of the solar corona and the driving mechanisms of solar wind are key challenges in understanding solar physics. A full three-dimensional compressible magnetohydrodynamic (MHD) simulation was conducted to distinguish between the heating mechanisms in the fast solar wind above the open field region.
Our simulation describes the evolution of the Alfv\'{e}nic waves, which includes the compressible effects from the photosphere to the heliospheric distance $s$ of 27 solar radii ($R_\odot$).
The hot corona and fast solar wind were reproduced simultaneously due to the dissipation of the Alfv\'{e}n waves.
The inclusion of the transition region and lower atmosphere enabled us to derive the solar mass loss rate for the first time by performing a full three-dimensional compressible MHD simulation.
The Alfv\'{e}n turbulence was determined to be the dominant heating mechanism in the solar wind acceleration region ($s>1.3 R_\odot$), as suggested by previous solar wind models. In addition, shock formation and phase mixing are important below the lower transition region ($s<1.03R_\odot$) as well. 
\end{abstract}




\section{Introduction}

Deciphering the heating mechanisms of the solar corona and the driving mechanisms of solar wind have been long standing problems in solar physics.
Both problems are deeply coupled with each other and they are involved with the solar mass loss process that can be applicable to solar-like stars.
For example, \cite{2011ApJ...741...54C} applied their solar wind model to predict the mass loss rates of cool stars.
In addition, \cite{2013PASJ...65...98S} suggested a theoretical interpretation for the saturation problem of stellar wind \citep{2005ApJ...628L.143W}.
Moreover, the angular-momentum loss rate has been estimated to yield a spin down law of the stars \citep{2020ApJ...896..123S}.
The purpose of this study is to provide a deep insight into the physics behind mass loss processes, while focusing on the Sun as one of the ubiquitous stars in the universe. 

It is widely accepted that photospheric convection is the energy source for the hot corona and solar wind \citep{2006SoPh..234...41K}.
One of the most plausible energy carriers  is the magnetohydrodynamic (MHD) wave, which is excited through the interactions between the convection and flux tubes that are embedded there. 
Among the MHD waves, the Alfv\'{e}n waves can carry energy to the corona 
because the slow waves will be dissipated down in the chromosphere and the fast waves will be refracted before they reach the corona \citep{2002AdSpR..30..471D}.
To the lowest order, only the Alfv\'{e}n waves will not suffer dissipation or refraction to transport the wave energy upward to the corona.

Numerous competing models have been proposed to dissipate the kinetic and magnetic energy in the Alfv\'{e}n waves in the corona \citep[e.g.][]{2000ApJ...530..999M}.
The Alfv\'{e}n waves can be steepened into a train of weak shocks \citep{1982ApJ...254..806H,2000A&A...353..741N}, which results in coronal heating.
Besides the shock heating, phase mixing is also suggested as a damping mechanism of the Alfv\'{e}n waves when there is in-homogeneity in the Alfv\'{e}n speed across the magnetic field lines \citep{1983A&A...117..220H}.
Moreover, the Alfv\'{e}n wave turbulence that is driven by the collisions between the counter propagating Alfv\'{e}n waves will work efficiently as a coronal heating mechanism \citep{1999ApJ...523L..93M}.
However, the dominant process in the solar atmosphere among these competing models has not been elucidated.

\vspace{\baselineskip}  

There have been a variety of attempts to theoretically reproduce the coronal and solar wind structure through Alfv\'{e}n wave dissipation.
\cite{2005ApJ...632L..49S} suggested a shock heated solar wind model where the hot corona and fast solar wind are reproduced as a natural consequence of the Alfv\'{e}n wave injection from the photosphere.
Because their models are based on one-dimensional MHD simulation, the contribution from the Alfv\'{e}n turbulence is neglected.
Alfv\'{e}n wave turbulence has been adopted as one of the heating mechanisms in the other models to self-consistently reproduce the corona and solar wind \citep{1986JGR....91.4111H,2007ApJS..171..520C,2010ApJ...708L.116V,2011ApJ...743..197C,2018ApJ...853..190S}.
Although some of these models include the shock effect and the turbulence simultaneously, simplified formulations are used to derive the heating rate without solving the full dynamics of the Alfv\'{e}n waves. 
The recent development of computational power enables us to survey the full dynamics of the Alfv\'{e}n waves over the global solar wind structure \citep{2016ApJ...821..106V,2019ApJ...880L...2S,2019JPlPh..85d9009C}.
However, so far there is no self-consistent simulation where the global solar wind structure and Alfv\'{e}n wave dynamics are solved simultaneously.

In this study, we perform full three-dimensional (3D) MHD simulations to reproduce the corona and solar wind through the Alfv\'{e}n wave scenario.
The basic strategy is similar to that of the previous studies \citep{2005ApJ...632L..49S,2006JGRA..11106101S,2012ApJ...749....8M,2014MNRAS.440..971M}.
We focused on a single flux tube that is extended from the polar regions on the quiet Sun.
Subsequently, the flux tube was perturbed in the photosphere to inject the Alfv\'{e}n waves into the upper atmospheres.
Next, we continued our numerical simulation until the system reached a quasi-steady state where radiative, adiabatic, and conductive cooling were balanced with the Alfv\'{e}n wave dissipation.
With this approach, the heating rate and background atmosphere were self-consistently determined simultaneously.

\section{Models and Assumptions}

In the present study, we attempted to mimic a single flux tube that is extended from a kilo-Gauss patch in the polar region \citep{2008ApJ...688.1374T,2010ApJ...719..131I,2012ApJ...753..157S} by using fully compressible 3D MHD simulations that include the effects of thermal conduction and radiative cooling. 
\begin{eqnarray}
 \frac{\upartial \rho}{\upartial t} &+& \nabla \cdot \left( \rho \mathbfit{v} \right) = 0,\\
 \frac{\upartial \rho \mathbfit{v}}{\upartial t} &+& \nabla \cdot \left( P_{\rm g} + \frac{B^2}{2} + \rho \mathbfit{vv} - \mathbfit{BB} \right) = \rho \mathbfit{g} + \mathbfit{F}_{\rm ex}, \\
 \frac{\upartial \mathbfit{B}}{\upartial t} &+& \nabla \cdot \left( \mathbfit{vB} - \mathbfit{Bv}\right) = 0,\\
 \frac{\upartial \cal E}{\upartial t} &+& \nabla \left[ \left({\cal E} + P_{\rm g} + \frac{B^2}{2} \right) \mathbfit{v} - \left( \mathbfit{B}\cdot\mathbfit{v} \right) \mathbfit{B} \right] \\
 &=& \rho \mathbfit{v}\cdot\mathbfit{g} + \nabla \cdot \mathbfit{q} + Q_{\rm rad} + \mathbfit{F}_{\rm ex}\cdot \mathbfit{v}, 
\end{eqnarray}
where ${\cal E}$ is the total energy density, 
\begin{eqnarray}
  {\cal E} &=& \frac{1}{2}\rho v^2 + e_{\rm int} + \frac{1}{2} B^2, 
\end{eqnarray}
$\rho,\mathbfit{v},\mathbfit{B},e_{\rm int},$ and $P_g$ are the mass density, fluid velocity, the magnetic field normalized against $\sqrt{4\upi}$, internal energy, and the gas pressure, respectively. An approximated equation of the state was used in our simulation, while assuming the mean molecular weight is a function of the temperature \citep{2014MNRAS.440..971M}.
Here, we did not include any explicit dissipation terms in the basic equations.
Therefore, all the heating terms are taken from the numerical dissipation.
The reliability of this numerical dissipation  is discussed in section \ref{sec:results}.

We assumed a slender flux tube with a width of 6 Mm for the photosphere and 27 R$_\odot$ (18,900 Mm) in length.
We adopted a modified spherical coordinate system, $(s,\theta,\phi)$, with the metric $\mathsf{g}$, which includes the effect of the super-radial expansion due to the dipole field of the quiet sun.
\begin{eqnarray}
 \mathsf{g}=\left(
 \begin{array}{ccc}
  1 & 0   & 0 \\
  0 & r(s)^2 & 0 \\
  0 & 0   & r(s)^2 \\
 \end{array} 
 \right)
 \end{eqnarray}
where $s$ is heliospheric distance and $\theta,\phi$ are the same angles in the standard spherical coordinate system.
The entire numerical domain is covered by $(N_s,N_\theta,N_\phi)=(6500,90,90)$ grids.
The grid size is uniform in the $\theta$ and $\phi$ directions, while the radial grid size increases from 25 km to 3 Mm to cover the entire radial domain ($1-27~R\odot$).

Because we assumed a slender flux tube ($ 6~{\rm Mm} << R_\odot$), the $\theta-\theta$ and $\phi-\phi$ components are equivalent in the metric. The radius of the numerical domain $r(s)=s\sqrt{f(s)}$ was determined by the same analytical super-radial expansion factor that was used in \cite{1976SoPh...49...43K}:
\begin{eqnarray}
 f(s) = \frac{f_{\rm max}\exp\left(\frac{s-s_1}{\sigma_1}\right)+f_1 }{\exp\left(\frac{s-s_1}{\sigma_1}\right)+1},
\end{eqnarray}
where 
\begin{eqnarray}
 f_1=1-(f_{\rm max}-1)\exp \left( \frac{R_\odot-s_1}{\sigma_1}\right).
\end{eqnarray}
In this study, we set $f_{\rm max}=3$, $\sigma_1=0.1$ $R_\odot$, and $s_1=1.2$ $R_\odot$, which mimics the super-radial expansion above the coronal hole due to the large scale dipolar structure during the quiet sun.
If $f(s)$ is set to 1, the modified spherical coordinate system is reduced to the standard spherical coordinate system.

Thermal conduction was assumed to have Spitzer-type conductivity that has 
a strong temperature dependence and it is only efficient along the magnetic field line.
The heat flux, $\mathbfit{q}$, can be written as 
\begin{eqnarray}
 \mathbfit{q}=-\kappa_0 \xi(r) T^{5/2} \frac{\mathbf{B}\cdot\nabla T }{B^2}\mathbf{B},
\end{eqnarray}
where $\kappa_0$ is $10^{-6}$ in cgs and $\xi(r)$ is the same quenching term that was introduced by \cite{2019ApJ...880L...2S} to approximately include the effect of collision less nature above $s>5R_\odot$.

Two different cooling functions were combined in our simulation to describe the radiative cooling in the optically thick layer and the optically thin layer \citep{2007ApJS..171..520C}:
\begin{eqnarray}
 Q_{\rm rad} = e^{-\tau_R/\tau_0}Q_{\rm thin} + (1-e^{-\tau_R/\tau_0}) Q_{\rm thick},
\end{eqnarray}
where $\tau_0=0.1$ and $\tau_R$ represent the Rossland mean optical depth. 
In our simulation, $\tau_R$ was assumed to be a function of the height for simplicity and it is written as
\begin{eqnarray}
  \tau_R = e^{-(s-R_\odot)/H_R}
\end{eqnarray}
where $H_R=155$ km in this study.
For the cooling in the optically thin layer, $Q_{\rm thin}$, we adopted the same polynomial fit \citep{2014MNRAS.440..971M} to the 
cooling function \citep{1990A&AS...82..229L}.
For the cooling in the optically thick layer $Q_{\rm thick}$, we used the same functions in \citep{2007ApJS..171..520C}, while assuming
local thermal equilibrium and a grey atmosphere.

\begin{figure*}
	\includegraphics[width=160 mm]{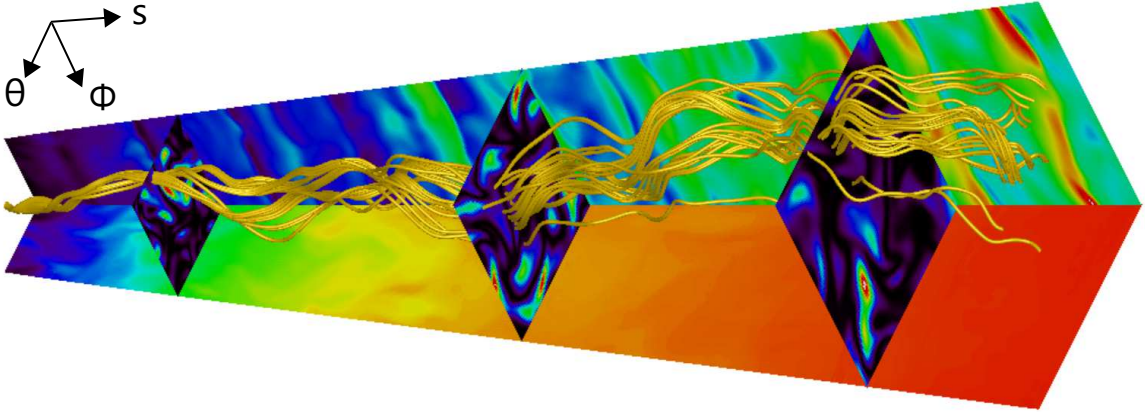}
    \caption{External view of our numerical model in the quasi-steady-state.
    The temperature structure (on $\phi=0$) and velocity structure (on $\theta=0$) are projected on the bottom and side surfaces, respectively.
    The horizontal slices indicate the structures of square current densities (on $s=2,3,4 R_\odot$).
    The yellow lines represent the magnetic field lines.
    The aspect ratio has been changed to emphasize the global structure.
    }
    \label{fig:3dview}
\end{figure*}

To mimic the buffeting motion of the flux tubes, we employed a given volumetric force that was localized at the foot point of the flux tube.
\begin{eqnarray}
 \mathbfit{F}_{\rm ex} = \mathbfit{F}(t) \exp{\left[  -\frac{1}{2} \left( \frac{s-R_\odot-\Delta s}{w_\parallel} \right)^2 - \frac{R_\odot^2}{2} \left( \frac{\theta^2+\phi^2}{w_\perp^2} \right) \right]},
\end{eqnarray}
where $\mathbfit{F}(t)$ was the amplitude of the external force with only the $\theta$ and $\phi$ components that had a white noise spectrum in a given frequency range ($[10^{-3},2\times 10^{-2}]$ Hz).
$w_\perp$ (=400 km) and $w_\parallel $ (=100 km) determined the size of the enforced region.
$\Delta s$ (=400 km) gave the height of the external force.
The amplitude of the external force was adjusted so that the root mean square (r.m.s.) of the resultant transverse velocity for the photosphere became $\sim$ 1 km/s.
This is consistent with the observed photospheric convection motion \citep{2010ApJ...716L..19M,2012ApJ...752...48C,2020ApJ...890..141O}. 

We adopted a potential field for the initial magnetic field.
The potential field was extrapolated from the bottom boundary with
\begin{eqnarray}
 B_s(s=R_\odot,\theta,\phi) = B_0 \exp\left(-R_\odot^2 \frac{\theta^2+\phi^2}{w_B^2}  \right) + \delta B,
\end{eqnarray}
where $w_B=700$ km and $B_0=1,800$ G.
The constant term $\delta B$ ($\sim-70$ G) was determined so that the mean radial magnetic field became $6$ G in the photosphere.

The initial atmospheric structures had a combination of the hydrostatic atmosphere and isothermal solar wind structure. 
For the hydrostatic atmosphere, we adopted the temperature distribution in the standard solar atmosphere, such as the VAL-C model \citep{1981ApJS...45..635V} with a bottom mass density of 10$^{-7}$ g cm$^{-3}$.
The isothermal wind \citep{1958ApJ...128..664P} with 1.1 MK was smoothly connected to the hydrostatic atmosphere at $s = 1.14~R_\odot$.
Because the initial condition did not satisfy the energy balance, the system will evolve to reach a new quasi-steady state.

At the outer end of the boundary, we posed free boundary conditions where we assumed a zero gradient for all the conservative variables.
For the inner boundary, the mass density and internal energy were extrapolated logarithmically.
All the components of the velocity were set to be zero while the magnetic field was extrapolated with a fourth order of accuracy. 
We posed periodic boundary conditions in the $\theta$ and $\phi$ directions.

We developed an original MHD code to perform accurate and stable simulations even in the low beta region around the transition region.
We adopted the HLL approximated Riemann solver \citep{1991JCoPh..92..273E} for the numerical flux.
The conservative variables were reconstructed in each cell by using a third order weighted essentially non-oscillatory (WENO) scheme. This was then integrated according to time by a third order arbitrary derivative Riemann problem scheme \citep{2009JCoPh.228.2480B}.
The divergence-free reconstruction allowed us to reduce $\nabla \cdot \mathbfit{B}$ to truncation errors \citep{2009JCoPh.228.5040B}.
Because the time scale of the thermal conduction is generally much shorter than the dynamics, we adopted an operator split method and we implicitly solved the thermal conduction by using the super time stepping method \citep{2012MNRAS.422.2102M}.

\section{Results } \label{sec:results}

\begin{figure}
	\includegraphics[width=87 mm]{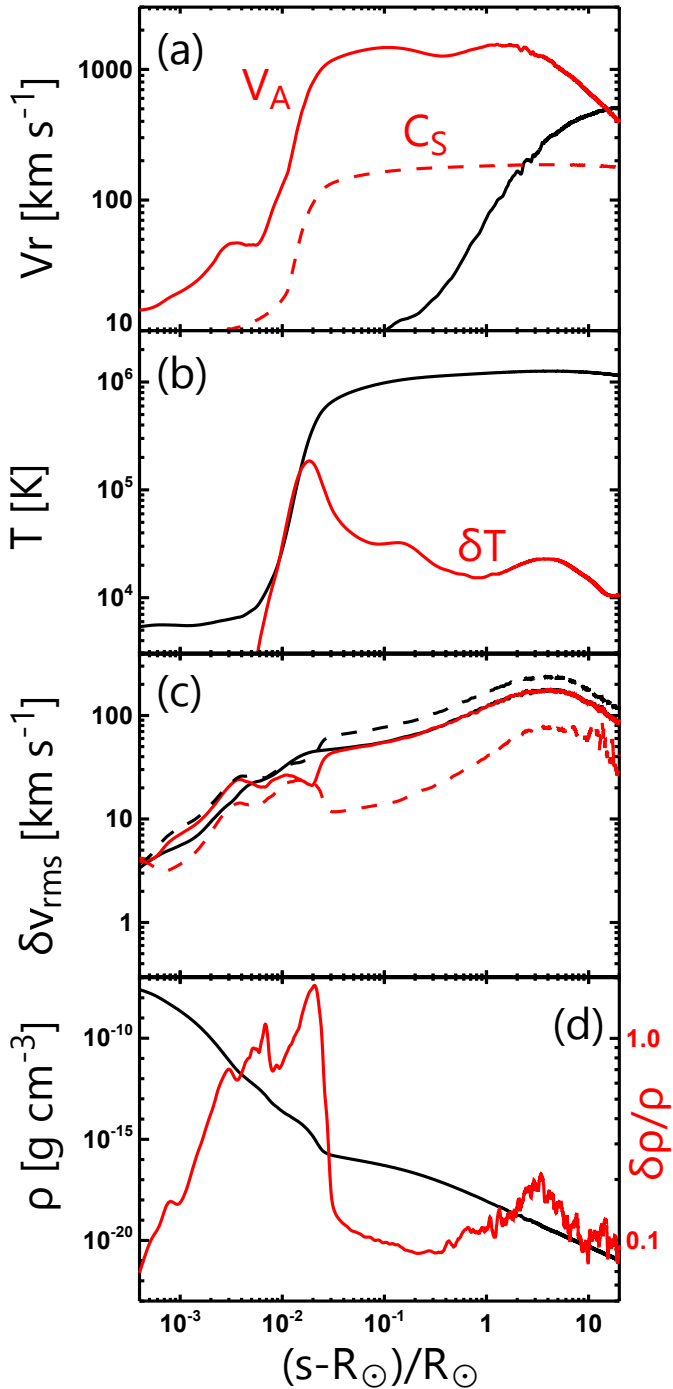}
    \caption{Global structures of our numerical model in the quasi-steady state:
    (a) Radial velocity (black line), the Alfv\'{e}n speed (red solid line), and the sound speed (red dashed line).
    (b) Temperature (black line) and the r.m.s. of the temperature fluctuation (red line).
    (c) The r.m.s. of the horizontal velocities $v_{\perp;{\rm rms}}$ (black solid line), magnetic field fluctuations $B_{\perp;{\rm rms}}/\sqrt{\langle \rho \rangle}$ (red solid line), the outgoing Els\"{a}sser variables $Z^+_{\perp;{\rm rms}}/2$ (black dashed line), and the incoming Els\"{a}sser variables $Z^-_{\perp;{\rm rms}}/2$ (red dashed line).
    (d) Density (black line) and the r.m.s. of the fractional density $\delta \rho_{\rm rms}/ \langle \rho \rangle $ (red line).
    All the values are averaged across the flux tube and over time (125 min).
    }
    \label{fig:global_structure}
\end{figure}

After a few Alfv\'{e}n crossing time passed, the system reached a quasi-steady state that had both the hot corona and fast solar wind. 
The quasi-steady state was archived because the radiative, conductive, and adiabatic loss was balanced with the dissipation of the MHD waves, which was continuously excited in the photosphere. 
Fig. \ref{fig:3dview} shows the external view of the global structure ($1.5~R_\odot < s < 4.5~R\odot$) in our model.
The resultant mass loss rate was $2.4\times10^{-14}~{\rm M}_\odot~{\rm yr}^{-1}$, which is the typical solar mass loss rate. 
Fig. \ref{fig:global_structure} shows the global atmospheric structure of our numerical model. 
The flow speed is indicated by the black solid line in Fig. \ref{fig:global_structure}(a) and it exceeded the local sound (Alfv\'{e}n) speed at $s=$ 3.3 (16.2)$R_\odot$ and it reached 530 km s$^{-1}$ at the outer end of our simulation (27 $R_\odot$).
The maximum temperature of 1.3 MK was archived at $s= 5.3$  $R_\odot$.
The r.m.s. of the transverse velocity, $v_{\perp;{\rm rms}}$, is indicated by the black solid line in Fig. \ref{fig:global_structure}(c) and it increased from 3 km s$^{-1}$ to 180 km s$^{-1}$ at $s=5.3 R_\odot$, and then it decreased with $s$.
These bulk properties revealed that our model was consistent with the solar corona and the fast solar wind above the coronal holes \citep{2009LRSP....6....3C}.

The red solid line in Fig. \ref{fig:global_structure}(c) denotes the normalized magnetic fluctuation $B_{\perp {\rm ; rms}}/\sqrt{\langle \rho \rangle}$ that revealed the energy equipartition between $v_\perp$ and $B_\perp$ in $s>1.03 R_\odot$.
The bracket $<>$ denotes the averaging operator over the cross-section and time.
The equipartition was violated in $s<1.03 R_\odot$ due to the non-WKB effects \citep{2005ApJS..156..265C}.

An enhancement of the density fluctuations of about 10-20\% was found in the solar wind acceleration region ($s \sim 4R_\odot$), which may increase the reflection of the Alfv\'{e}n waves (the red line in Fig. \ref{fig:global_structure}d). 
We estimated the reflection rate while using the formulation in \cite{1980JGR....85.1311H} and determined that the reflection rate due to the fluctuation could be up to 50 times larger than the expected rate from the gradual change in the background Alfv\'{e}n speed.
This property has already been found in the literature \citep{2006JGRA..11106101S,2019ApJ...880L...2S}, which suggests that these  density fluctuations are generated through the parametric decay of the Alfv\'{e}n waves \citep{1969npt..book.....S,1978ApJ...219..700G}.
The obtained amplitude of 10-20 \% for the fractional density is 
consistent with the radio scintillation observation \citep{2014ApJ...797...51M}.

\begin{figure}
	\includegraphics[width=\columnwidth]{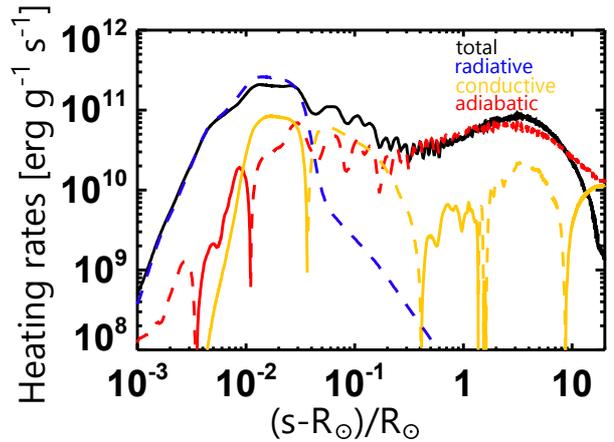}
    \caption{Heating and cooling rate as a function of the heliocentric distance.
    The blue, red, and yellow lines correspond to the radiative, adiabatic, and thermal conduction loss, respectively.
    The positive values (net heating) are indicated by the solid lines while the negative values (net cooling) are indicated by the dashed lines.
    }
    \label{fig:energy1d}
\end{figure}

The total mechanical heating rate per unit mass $Q_{\rm tot}$ was estimated by summing up all the cooling rates: adiabatic, radiative, and conduction cooling. 
When deriving the total heating rate, we assumed that the system was in the quasi-steady state and the energy balance was established statistically.
The black solid line in Fig. \ref{fig:energy1d} indicates the estimated total heating rate. 
The heating rate was averaged over the cross-section over 125 min. 
Because we did not include the explicit dissipation terms in the basic equations, all the heating came from the numerical dissipation. 
The derived heating rate included the contributions from all the mechanical heating mechanisms. Therefore, a further analysis is needed to distinguish the dominant heating mechanisms in our simulation.

The dominant cooling mechanisms changed with the heliospheric distance.
The blue, red, and yellow lines in Fig. \ref{fig:energy1d} correspond to the radiative, adiabatic, and thermal conduction loss, respectively.
The positive values (net heating) are indicated by the solid lines while the negative values (net cooling) are indicated by the dashed lines.
Below the chromosphere ($s<1.01 R_\odot$), the radiative cooling was dominant.
In the lower transition region ($1.01R_\odot< s<1.03 R_\odot$), the radiative cooling was the main cooling term whereas the thermal conduction mostly acted as a heating term in addition to the mechanical heating.
In the upper transition region ($1.03 R_\odot < s < 1.3~R_\odot$), the energy loss due to the thermal conduction balanced with the mechanical heating.
In the solar wind acceleration region ($s>1.3~R_\odot$), the plasma was adiabatically cooled down.

Besides the total heating rate, $Q_{\rm tot}$, we also estimated the numerical dissipation that was discovered to have a hyper diffusive nature.
According to the test simulations in the Appendix, the heating rates per unit mass can be estimated by
\begin{eqnarray}
 Q_{v;ij}&=&\nu_h \left( \frac{\upartial ^2 v_i}{\upartial x_j^2}  \right)^2 \\
 Q_{B;ij}&=&\frac{\nu_h}{\langle \rho \rangle} \left( \frac{\upartial ^2 B_i}{\upartial x_j^2}  \right)^2,
\end{eqnarray}
where $Q_{v;ij}$ and $Q_{B;ij}$ can be regarded as the dissipation rate of the kinetic and magnetic energy.
The subscripts $i,j$ denote the direction.
The hyper diffusivity was estimated to be
\begin{eqnarray}
 \nu_h = 0.07 (\Delta_i)^3 V_f,
\end{eqnarray}
where $\Delta_i$ is the grid size in the $i$ direction and $V_f$ is the local fast mode speed.
We refer to the dissipation of the field-aligned current and the vorticity to $Q_\parallel$ where
\begin{eqnarray}
 Q_\parallel \equiv Q_{v;\theta \phi} + Q_{B;\theta \phi} + Q_{v;\phi \theta}+Q_{B;\phi \theta}.
\end{eqnarray}
The heating rates may include the dissipation through the Alfv\'{e}n turbulence and phase mixing because they are related to the shears across the field lines.
The heating rate due to the dissipation of the perpendicular current and the vorticity $Q_\perp$ may be described as
\begin{eqnarray}
 Q_\perp \equiv Q_{v:\theta s} + Q_{v:\phi s} + Q_{B:\theta s} + Q_{B:\phi s}, 
\end{eqnarray}
which includes the shock heating rate because the nonlinear steepening of the Alfv\'{e}n waves will produce a discontinuous plane that is perpendicular to the field lines. 
We can use these two heating rates to distinguish the competing heating mechanisms that happened in the simulation.

\begin{figure}
	\includegraphics[width=\columnwidth]{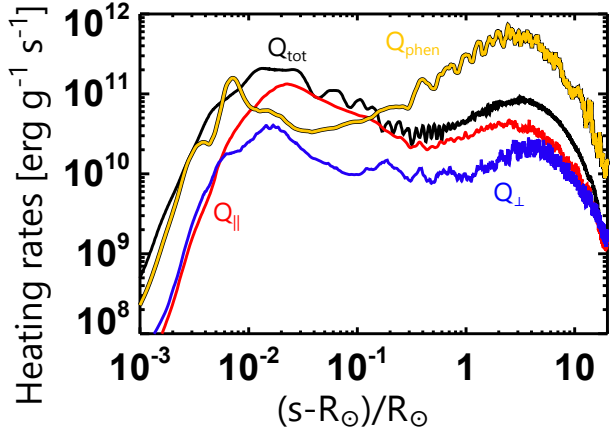}
    \caption{Dissipation rate of the kinetic and magnetic energy while assuming hyper diffusion.
    The black line indicates the total heating rate $Q_{\rm tot}$ and it is the same as the black line in Fig. \ref{fig:energy1d}.
    The red and blue lines are $Q_\parallel$ and $Q_\perp$, respectively.
    The yellow line denotes the phenomenological heating rate $Q_{\rm phen}$.
    }
    \label{fig:dissipation}
\end{figure}

The dominant heating process was determined to be the dissipation of the field aligned current, $Q_\parallel$, in $s>1.01R_\odot$.
Fig. \ref{fig:dissipation} compares $Q_{\rm tot}$, $Q_{\parallel}$, and $Q_\perp$ as a function of the heliocentric distance by using the 
 black, red, and blue lines, respectively.
The summation of $Q_\parallel$ and $Q_\perp$ was consistent with $Q_{\rm tot}$ within a factor of two  at $s>1.01R_\odot$. This also 
supports that our derivation of the numerical heating rate is valid in this region.
The contribution from $Q_\parallel$ was determined to be dominant in this region.
The amplitude of $Q_\parallel$ was roughly consistent with that of the turbulent heating rates reported in previous studies \citep{2002ApJ...575..571D,2007ApJS..171..520C,2016ApJ...821..106V}.
In $s<1.005 R_\odot$, $Q_\perp$ was comparable to $Q_\parallel$, which indicates that the shock heating is also important in this layer
\citep{1982SoPh...75...35H,1999ApJ...514..493K,2005ApJ...632L..49S,2010ApJ...710.1857M,2020ApJ...900..120S}.
However, the heating rate analysis must be less reliable because $Q_{\parallel}+Q_{\perp}$ is $\sim 4$ times smaller than $Q_{\rm tot}$.
This may be due to the switching of the reconstruction method in our numerical solver from WENO to minmod, which happened more frequently below the transition region to change the hyperdiffusive nature of our numerical solver.

\begin{figure}
	\includegraphics[width=\columnwidth]{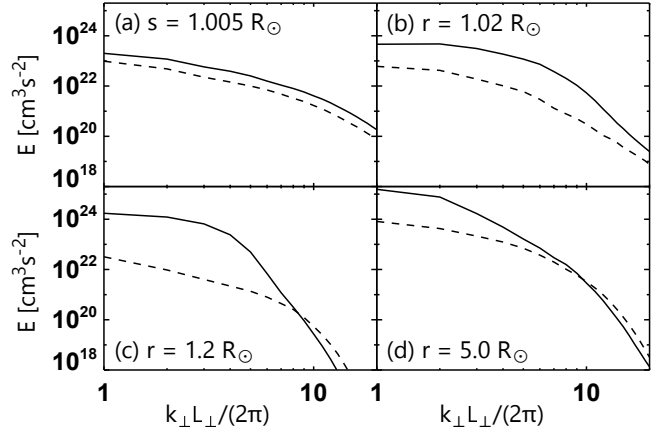}
    \caption{
    Estimated PSD of the Els\"{a}sser variables that are perpendicular to the mean magnetic field $E^\pm(k_\perp)$ that are measured at different heights:
    (a) $s=1.005R_\odot$, (b) $s=1.02 R_\odot$, (c) $s=1.2 R_\odot$ and (d) $s=5R_\odot$.
    The solid and dashed lines denote $E^+(k_\perp)$ and $E^-(k_\perp)$, respectively.
    }
    \label{fig:elsasser_spectra}
\end{figure}

The dynamics of the Alfv\'{e}n waves in the solar wind follow the typical behaviour in the MHD turbulence that is driven by the collisions between the counter propagating Alfv\'{e}n waves \citep{1965PhFl....8.1385K}.
This argument was demonstrated from the properties of the fine scale structures in the Els\"{a}sser variables \citep{1950PhRv...79..183E}.
\begin{eqnarray}
 \mathbf{Z}^\pm \equiv \mathbf{v} \mp \frac{\mathbf{B}}{\sqrt{\rho}},
\end{eqnarray}
where $\mathbf{Z}^+$ and $\mathbf{Z}^-$ represent the amplitude of the outward and inward propagating Alfv\'{e}n waves, respectively.
Fig. \ref{fig:elsasser_spectra} shows the estimated power spectral densities (PSD) of the Els\"{a}sser variables that are perpendicular to the mean magnetic field $E^\pm(k_\perp)$ at different heights where
\begin{eqnarray}
 \int E^\pm (k_\perp) dk_\perp = \frac{1}{\Omega} \int  Z_\perp^\pm(\theta,\phi)^2 d\theta d\phi
\end{eqnarray}
and $\Omega$ is the solid angle for the cross-section of the flux tube.
Fig. \ref{fig:elsasser_spectra}(d) indicates that the $E^-(k_\perp)$ is much flatter than $E^+(k_\perp)$.
Therefore, the finer structures tended to survive in $\mathbf{Z}^-_\perp$ than in $\mathbf{Z}^+_\perp$, which is consistent with the imbalanced turbulence where the amplitude of the outgoing waves is much larger than the incoming waves \citep{1983A&A...126...51G,2009PhRvL.103v5001B}. 
This is supporting evidence that the dynamics in our simulation are controlled by the Alfv\'{e}n wave turbulence.

\begin{figure}
	\includegraphics[width=\columnwidth]{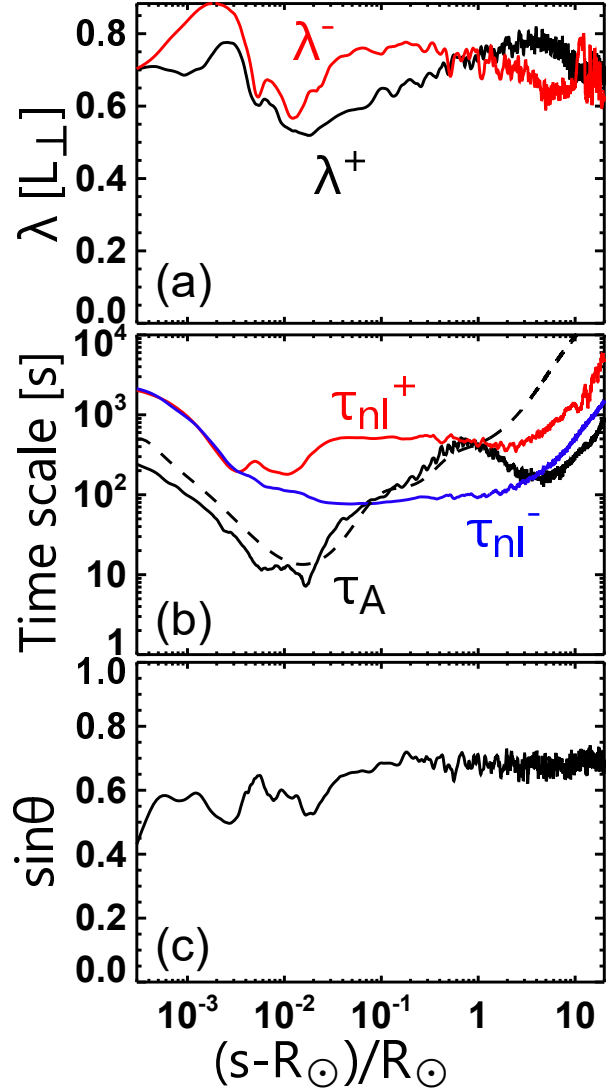}
    \caption{Properties of the MHD turbulence in regard to the radial distance:
    (a) Energy containing scale $\lambda^+$ (black line) and $\lambda^-$ (red line).
    (b) Nonlinear cascading time scale $\tau_{\rm nl}^+$ (red line), $\tau_{\rm nl}^-$ (blue line), and Alfv\'{e}n (reflection) time scale $\tau_A$ (black solid line).
    The dashed line indicates the Alfv\'{e}n time scale that was estimated from the smoothed background.
    (c) Sine of the alignment angle between $\mathbf{Z}_\perp^\pm$.
    }
    \label{fig:turb_prop}
\end{figure}

The phenomenological turbulent heating rate was frequently used in the solar wind model, which may need some modifications.
The yellow line in Fig. \ref{fig:dissipation} indicates the simplest phenomenological turbulent heating rate
\citep{1995PhFl....7.2886H,1999ApJ...523L..93M}, which is described as follows.
\begin{eqnarray}
 Q_{\rm phen} = \frac{(Z_\perp^+)^2 Z_\perp^-}{4\lambda^+} 
 + \frac{(Z_\perp^-)^2 Z_\perp^+}
 {4 \lambda^-},
\end{eqnarray}
where $Z_\perp^\pm$ is the r.m.s. amplitude of $\mathbf{Z}_\perp^\pm$ and the energy containing scale $\lambda^\pm$ was approximated by the width of the simulation box $L_\perp$.
The heating rate is enhanced when the energy containing scale of the Alfv\'{e}n turbulence becomes shorter. 
The energy containing scale may be estimated as follows.
\begin{eqnarray}
 \lambda^\pm \equiv \frac{2 \pi}{(Z^\pm_\perp)^2} \int dk_\perp \frac{E^\pm (k_\perp)}{k_\perp}
\end{eqnarray}
\citep{1995PhFl....7.2886H}.
The averaged value of $\lambda^\pm$ was $0.5 \sim 0.9 $ $L_\perp$ all over the numerical domain, which will increase $Q_{\rm phen}$ by at most over a factor of two (Fig. \ref{fig:turb_prop}a).
On the other hand, $Q_{\rm phen}$ may be reduced when the Alfv\'{e}n propagating time scale $\tau_A\sim L_\parallel/V_A$ is comparable or is shorter than the nonlinear cascading time scale $\tau _{\rm nl}^\pm =L_\perp /Z_\perp^\mp$ \citep{1980PhRvL..45..144D}.
The timescale ordering is essential for defining the efficiency of the Alfv\'{e}n turbulence \citep{2003ApJ...597.1097D}.
The approximated quenching factor is suggested \citep{2006PhPl...13d2306O,2007ApJS..171..520C}
to be
\begin{eqnarray}
 \frac{1}{1+\tau_{\rm nl}^\pm/\tau_A}.
\end{eqnarray}
If we estimated $\tau_A$ with $|\nabla \cdot V_A|^{-1}$, $\tau _A$ is almost comparable to or is smaller than $\tau _{\rm nl}^+$ (Fig. \ref{fig:turb_prop}b).
Especially at the transition region ($s\sim1.03R_\odot$), $\tau_A$ is 50 times smaller than $\tau  _{\rm nl}^+$, which may reduce the heating rate by 50 times.
On the other hand, $\tau_{\rm nl}^-$ was smaller than $\tau_A$ at $s$ $\sim$ 2 $R_\odot$.
The alignment angle $\theta$ between $\mathbf{Z}_\perp^\pm$ may be an important factor to modify $Q_{\rm phen}$ because the strength of the nonlinear coupling between the counter propagating Alfv\'{e}n waves is proportional to $|\mathbf{Z}_\perp^+ \times \mathbf{Z}_\perp^-|$ \citep{2013PhPl...20g2302H}.
The sine of the alignment angle was estimated as follows.
\begin{eqnarray}
 \sin \theta = \frac{\langle |\mathbf{Z}^+_\perp \times \mathbf{Z}^-_\perp |\rangle}{\langle |\mathbf{Z}_\perp^+| \rangle \langle |\mathbf{Z}_\perp^- |\rangle}.
\end{eqnarray}
Fig. \ref{fig:turb_prop}c indicates that the sine of the alignment angle was 0.5 to 0.7 throughout the domain. This is larger than what was estimated in the recent reduced MHD simulations \citep{2013ApJ...776..124P,2019JPlPh..85d9009C}. 
This discrepancy may arise as a result of the horizontal variation of the Alfv\'{e}n speed due to the density fluctuation.
When Alfv\'{e}n waves propagate in the density fluctuation, the polarization angle of these Alfv\'{e}n waves will be scattered to randomize the alignment angle.

The dominant cascading process in the lower transition region ($s\sim1.02 R_\odot$) can be distinguished if we compare the growth rate of the phase mixing and the Alfv\'{e}n turbulence.
The nonlinear evolution of the Els\"{a}sser variables can be represented by
\begin{eqnarray}
 \left. \frac{\upartial \mathbf{Z}_\perp^\pm}{\upartial t} \right| _{\rm non linear}=-\mathbf{Z}^\mp \cdot \nabla \mathbf{Z}^\pm_\perp,
\end{eqnarray}
that are composed of three different terms.
\begin{eqnarray}-
 \left(
  \overline{\mathbf{Z}^\mp} +
  \delta \mathbf{Z}^\mp_\parallel+
  \delta \mathbf{Z}^\mp_\perp
 \right) \cdot \nabla \mathbf{Z}^\pm_\perp,
\end{eqnarray}
where the over bar denotes the average over the cross-section of the flux tube and $\delta$ represents the remaining fluctuations.
The first term denotes the Alfv\'{e}n wave propagation. 
The second term denotes the phase mixing term because it comes from the Alfv\'{e}n speed fluctuation that is perpendicular to the mean magnetic field. 
The last term denotes the Alfv\'{e}n turbulence.
Therefore, the growth rate of the phase mixing ($\gamma_{\rm PM}^\pm$) and the Alfv\'{e}n turbulence ($\gamma_{\rm AT}^\pm$) can be estimated as follows.
\begin{eqnarray}
 \gamma_{\rm PM}^\pm &\sim&
 \frac{|\delta \mathbf{Z}^\mp_\parallel \cdot \nabla \mathbf{Z}^\pm_\perp |}
 { Z_\perp^\pm },
 \end{eqnarray}
and
 \begin{eqnarray}
 \gamma_{\rm AT}^\pm &\sim&
 \frac{|\delta \mathbf{Z}^\mp_\perp \cdot \nabla \mathbf{Z}^\pm_\perp |}
 { Z_\perp^\pm },
\end{eqnarray}
respectively \citep{2018ApJ...859L..17S}.
With the approximated growth rate, we can roughly distinguish which process is dominant.

\begin{figure}
	\includegraphics[width=\columnwidth]{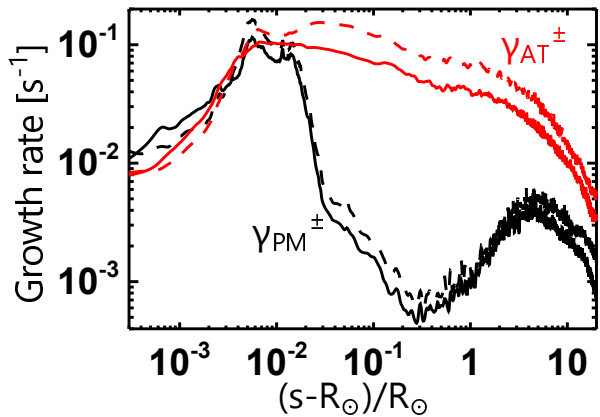}
    \caption{Time scales for phase mixing $\gamma _{\rm PM}^\pm$ (black lines) and the Alfv\'{e}n turbulence $\gamma_{\rm AT}^\pm$ (red lines) as a function of the heliocentric distance.
    The solid and dashed lines represent the plus and minus superscripts, respectively.
    }
    \label{fig:cmp_growth_rate}
\end{figure}

According to the analysis of the growth rates that are defined above, the dominant process in energy cascading changes with respect to the heliocentric distance. 
Fig. \ref{fig:cmp_growth_rate} shows the growth rates of the phase mixing and the Alfv\'{e}n turbulence.
The black and red lines represent $\gamma_{\rm PM}^\pm$ and $\gamma_{\rm AT}^\pm$, respectively.
The solid and dashed lines are represented by the plus and minus superscripts.
The growth rates of the phase mixing are comparable to those of the Alfv\'{e}n turbulence at $s<1.02R_\odot$. 
Above that, the Alfv\'{e}n turbulence dominated the cascading process.
Although the phase mixing started to be efficient around $s=$ 4 $R_\odot$ due to the density fluctuations from the parametric decay, the contribution was much smaller than the Alfv\'{e}n turbulence. 

In the upper transition region ($1.03 R_\odot < s < 1.3 R_\odot$), 
the resultant heating rate was overestimated owing to the numerical dissipation
until the steady Alfv\'{e}n turbulence was switched on above $s>1.3~ R_\odot$.
Because the efficiency of the Alfv\'{e}n turbulence becomes significantly small due to the lack of the reflection wave, the turbulent heating rate was not likely to sustain the conductive loss.
The analysis of the growth rate suggests that the phase mixing is also inactive in the upper transition region.
Therefore, we suggested that the dissipation of 
the turbulent structure
developed below the transition region was decaying to heat the upper transition region.
The decaying process can be seen in the turbulent spectra in Fig \ref{fig:elsasser_spectra}(b) and (c) where $E^+(k_\perp)$ was decaying without nonlinear cascading
, which indicates that this is not the turbulent heating.
Because the decaying process strongly depends on the diffusivity of the system, the heating here was most likely 
overestimated owing to the numerical diffusion. 
We note that the dissipation here is unrealistically large, and the full cascade is not accounted for.
The heating rate will reduce when the spatial resolution is increased.
The reduction of the heating rate may continue until it becomes smaller than the shock heating rate $\sim Q_\perp$, which may be independent from the spatial resolution.

The signature of energy cascading toward a smaller spatial scale across the flux tube was also found in the chromosphere, which is likely to contribute to the chromospheric heating.
Fig. \ref{fig:elsasser_spectra}(a) shows $E(k_\perp)^\pm$ at $s=1.005$ $R_\odot$.
The spectral shape of $E(k_\perp)^\pm$ in the chromosphere is similar to each other unlike the solar wind acceleration region where the outward propagating waves are dominant.
The cascading was possibly controlled by the phase mixing and the Alfv\'{e}n turbulence because $\gamma_{\rm PM} \sim \gamma_{\rm AT}$.

\section{Discussion and Conclusions}
We conducted a full compressible 3D MHD simulation of the corona and solar wind.
Our model succeeded in capturing the temperature and bulk speed in the corona and solar wind as a natural consequence of the continuous Alfv\'{e}n wave generation in the photosphere. 
The main dissipation mechanism for the solar wind acceleration region is the Alfv\'{e}n turbulence, although the compressibility played an important role in enhancing the reflection rate that controls the turbulent heating rate.
On the other hand, the dynamics in the chromosphere may not be described only by the Alfv\'{e}n turbulence. It is necessary to take phase mixing and the shocks into account.

This study describes the first solar wind model where the mass loss from the sun was self-consistently obtained by using the full compressible 3D MHD simulation.
This can be demonstrated only after we included the chromosphere at the bottom of the solar wind model using the 3D MHD simulation.
This simulation will be a milestone in applying the solar wind model to the other stars with the magnetically driven winds.

The dynamics of the Alfv\'{e}n waves in the solar wind acceleration region ($s>1.3 R_\odot $) are the same as those described in \cite{2019ApJ...880L...2S} even though we included the chromosphere. 
The dissipation of the Alfv\'{e}n waves was caused by the Alfv\'{e}n turbulence that was driven by counter propagating waves.
The dynamics came from the pure 3D nature of the Alfv\'{e}n wave collisions \citep{1965PhFl....8.1385K} and it cannot be described by 2D simulations \citep{2012ApJ...749....8M,2014MNRAS.440..971M}.
Such 3D collisions have been modelled based on the context of coronal heating by \cite{2003ApJ...597.1097D}.
Although the cascading process in the Alfv\'{e}n turbulence was incompressible, compressibility played an important role.
The outwardly propagating Alfv\'{e} waves suffered parametric decay \citep{1969npt..book.....S,1978ApJ...219..700G}, which generated inwardly propagating Alfv\'{e}n waves and density fluctuations.
The generated fluctuations were suggested to enhance the reflection rate that is more than 50 times larger than what was expected from the smooth change in the background Alfv\'{e}n speed \citep{2016ApJ...821..106V,2018ApJ...853..190S}.
The direct and indirect generation of inwardly propagating Alfv\'{e}n waves due to a compressible process like the parametric decay can enhance the turbulent heating rate by increasing the $Z^-_\perp$ components.
On the other hand, the efficiency of the turbulent heating rate may be reduced as the reflection time scale becomes smaller due to the compressible effects.

Although coronal heating mechanisms reported as per previous 1D \citep{2006JGRA..11106101S} and 2D \citep{2012ApJ...749....8M} simulations differ from those reported in our simulation, the basic properties of the wind, such as mass flux (or mass loss rate), do not differ significantly in our 3D simulation.
Further simulations are required to elucidate whether this coincidence is accidental or inevitable.
There may exist possibilities that 1D, 2D, and 3D simulations will show different dependences of the mass flux on velocity fluctuations at the photosphere or on the magnetic field structure.

The Alfv\'{e}n wave dynamics in the lower transition region and the chromosphere ($s<1.03 R_\odot$) are partially different from what can be described according to the theories of Alfv\'{e}n turbulence.
The previous studies investigated the effects of the Alfv\'{e}n turbulence with reduced MHD formulations that are below the transition regions \citep{2007ApJ...662..669V,2010ApJ...708L.116V,2011ApJ...736....3V,2012A&A...538A..70V} where the reduced  MHD approximation is marginally valid.
According to our analysis of the growth rate, the contribution to the energy cascading of the phase mixing is comparable to the Alfv\'{e}n turbulence in this region.
Therefore, the dynamics of the Alfv\'{e}n waves in this layer cannot be described only by the Alfv\'{e}n turbulence.
The effect of the phase mixing is small in the solar wind acceleration region even though it is enhanced by density fluctuations due to parametric decay.

The decaying of the turbulent structure was identified to heat the upper transition region ($1.03 R_\odot<s<1.3 R_\odot$), which probably overestimated the heating rate due to the lack of the spatial resolution.
The phase mixing and Alfv\'{e}n turbulence were inactive in terms of cascading wave energy.
The local heating rate may reduce when the spatial resolution is increased.
However, better resolution may also enhance the Poynting flux entering the corona because the wave dissipation in the chromosphere will reduce in that case.
In the context of the coronal loops, the coronal temperature and density can increase with the spatial resolution \citep{2016MNRAS.463..502M}.
If better resolutions lead to higher Poynting fluxes even above the open flux region, we speculate that the resultant solar wind would be hotter, heavier, and slower \citep{2006JGRA..11106101S}. 

There exist both advantages and disadvantages of direct simulations of solar wind.
One obvious advantage is that the involved physics do not need to be specified beforehand.
Because the dynamics of Alfv\'{e}n waves include nonlinear phenomena in inhomogeneous plasma, it is difficult to predict all the physical processes in advance.
For example, the importance of the parametric decay in enhancing the reflection rate has been clearly recognized only after the achievement of a successful direct simulation \citep{2006JGRA..11106101S,2019ApJ...880L...2S}.
Moreover, we found that the polarization angle of Alfv\'{e}n waves is likely to be scattered owing to the density fluctuation to maintain the polarization angle between counter-propagating Alfv\'{e}n waves.
The disadvantage of the direct simulations is the insufficiency of spatial resolution.
Because the magnetic Reynolds number is extremely large in general, the dissipation coefficients are overestimated in the direct simulations.
In other words, the resultantly obtained heating rate is sometimes unreliable, as in the case for the upper transition region in the present study. 
The models with subgrid scale physics \citep{2001ApJ...548..482D,2007ApJS..171..520C,2010ApJ...708L.116V} would have advantages in this context, because these models can capture the dynamics of Alfv\'{e}n waves well even in the case of plasma with a high magnetic Reynolds number.
However, the subgrid models sometimes overlook important physics aspects such as the coupling between the Alfv\'{e} wave and the density fluctuation. 
Nonetheless, if we can formulate the effects of compressibility, we can improve the subgrid models.

Although our model succeeded in creating the bulk properties of the corona and solar wind, there are some problems.
First, our model assumed only a single flux tube and it posed a periodic boundary in the horizontal direction. 
This assumption forces the dissipation process to occur inside the flux tube and it neglects the dissipation between the flux tubes. 
The interaction with the surrounding flux tubes or the emerging flux may cause a magnetic reconnection to feed the energy to the wind \citep{1999JGR...10419765F}.
Second, our model did not include the surface convection zone. 
Instead, we assumed that there was artificial translational forcing to generate the Alfv\'{e}nic waves in the upper photosphere.
This assumption ignores the torsional motions that can be expected at the intersections of the granular boundaries.
The torsional motions are considered to be efficient energy sources in the upper atmosphere and they were observed as magnetic tornadoes \citep{2012Natur.486..505W}.
Exclusion of the upper convection zones also forced our model to assume the magnetic field structure that is described above.
We assumed the potential field as the initial condition, where the force balance is determined only by the magnetic forces.
As time goes on, the force balance changes slightly. However, the magnetic field structure in the quasi-steady state was almost the same as the initial field because we posed fixed boundary conditions at the bottom.
In reality, the gas pressure also works in the force balance, which will change the property of the wave mode around the flux tube \citep{2005ApJ...631.1270H,2011ApJ...727...17F,2015ApJ...799....6M}.
Finally, our model focused on the fast solar wind above the polar region and it did not intend to reproduce the slow wind around the equatorial plane.
The slow solar wind may be demonstrated by changing the expansion factor of the flux tubes
\citep{2006ApJ...640L..75S}.
In future work, we will attempt to reproduce solar wind and compare the result with the observations from the Parker Solar Probes \citep{2019Natur.576..237B}.





\section*{Acknowledgements}
The author appreciates the anonymous referee for providing constructive
comments.
Numerical computations were carried out on Cray XC50 at the Center for Computational Astrophysics, at the National Astronomical Observatory of Japan. The numerical analyses were in part carried out on the analysis servers at the Center for Computational Astrophysics at the National Astronomical Observatory of Japan. A part of this study was carried out by using the computational resources from the Center for Integrated Data Science at the Institute for Space-Earth Environmental Research, Nagoya University through the joint research program. We would like to thank Editage (www.editage.com) for English language editing.
 
 \section*{Data Availability}
 The data underlying this article will be shared on reasonable request to the corresponding author.





\bibliographystyle{mn2e}
\bibliography{mn2e-jour,myref}




\appendix

\section{Hyperdiffusive nature in our numerical solver}
We performed 1D MHD simulations to confirm the numerical heating rates that can be roughly estimated through hyperdiffusion.
For the initial condition, we set the background magnetic field and the density to $\mathbf{B}_0 = \hat{x}$ and $\rho_0=1$, respectively. 
We changed the plasma beta ($\beta \equiv C_s^2/V_A^2$) from $0.01$ to $1$ as a parameter.
For the disturbed variables, 
\begin{eqnarray}
 v_y &=&  0.05 \sin \left( k z \right) \\
 B_y &=& -0.05 \sin \left( k z \right), 
\end{eqnarray}
where $k$ was a wave number in the $z$ direction.
Then, we performed 1D MHD simulations in $z$ direction with the periodic boundary condition by changing $\beta$ and $k$.
As time went on, the initial perturbations were dissipated through numerical diffusion and the magnetic and kinetic energy decreased exponentially with the diffusion time.

From the relationship between the measured diffusion time and the wave number $k$, the numerical dissipation was determined to act as the hyperdiffusion.
The measured diffusion time was proportional to $k^{-4}$, which is the property of the hyperdiffusion that has a fourth order derivative.
\begin{eqnarray}
 \frac{\upartial f}{\upartial t} = \nu_h \frac{\upartial^4 f}{\upartial z^4},
\end{eqnarray}
where $f$ is $v_y$ or $B_y$ in this case.
The hyperdiffusivity $\nu_h$ depends on the plasma beta and it can be fitted by
\begin{eqnarray}
 \nu_h \sim 0.07 \Delta_z^3 V_f.
\end{eqnarray}
where $V_f^2=V_A^2(1+\beta)$.

\bsp	
\label{lastpage}
\end{document}